\documentclass[preprint,
showpacs,preprintnumbers,
nofootinbib,
nobibnotes,
amsmath,amssymb,
aps,prd]{revtex4-1}
\usepackage{epsfig}
\usepackage{graphicx}
\usepackage{bm}
\usepackage{times}
\usepackage{braket}
\usepackage{color}
\usepackage{slashed}
\usepackage{hyperref}
\usepackage{ulem}
\definecolor{Red}{rgb}{1.,0.,0.}

\definecolor{Blue}{rgb}{0.,0.,1.}

\definecolor{Green}{rgb}{0.,1.,0.}

\definecolor{Gray}{rgb}{0.5,0.5,0.5}

\definecolor{nicered}{rgb}{0.7,0.1,0.1}
\definecolor{nicegreen}{rgb}{0.1,0.5,0.1}
\hypersetup{colorlinks,citecolor=nicegreen,linkcolor=nicered}

\begin{document}

\newcommand{\beq}{\begin{eqnarray}}
\newcommand{\eeq}{\end{eqnarray}}
\newcommand{\ben}{\begin{enumerate}}
\newcommand{\een}{\end{enumerate}}
\newcommand{\non}{\nonumber\\ }
\newcommand{\jpsi}{J/\Psi}
\newcommand{\ppa}{\phi_\pi^{\rm A}}
\newcommand{\ppp}{\phi_\pi^{\rm P}}
\newcommand{\ppt}{\phi_\pi^{\rm T}}
\newcommand{\ov}{ \overline }
\newcommand{\zerot}{ {\textbf 0_{\rm T}} }
\newcommand{\kt}{k_{\rm T} }
\newcommand{\fb}{f_{\rm B} }
\newcommand{\fk}{f_{\rm K} }
\newcommand{\rk}{r_{\rm K} }
\newcommand{\mb}{m_{\rm B} }
\newcommand{\mw}{m_{\rm W} }
\newcommand{\im}{{\rm Im} }
\newcommand{\kks}{K^{(*)}}
\newcommand{\acp}{{\cal A}_{\rm CP}}
\newcommand{\pb}{\phi_{\rm B}}
\newcommand{\xeba}{\bar{x}_2}
\newcommand{\xsba}{\bar{x}_3}
\newcommand{\peas}{\phi^A}
\newcommand{\Dsl}{ D \hspace{-2truemm}/ }
\newcommand{\pvsl}{ p \hspace{-2.0truemm}/_{K^*} }
\newcommand{\esl}{ \epsilon \hspace{-2.1truemm}/ }
\newcommand{\psl}{ p \hspace{-2truemm}/ }
\newcommand{\ksl}{ k \hspace{-2.2truemm}/ }
\newcommand{\lsl}{ l \hspace{-2.2truemm}/ }
\newcommand{\nsl}{ n \hspace{-2.2truemm}/ }
\newcommand{\vsl}{ v \hspace{-2.2truemm}/ }
\newcommand{\zsl}{ z \hspace{-2.2truemm}/ }
\newcommand{\epsl}{\epsilon \hspace{-1.8truemm}/\,  }
\newcommand{\bfkk}{{\bf k} }
\newcommand{\calm}{ {\cal M} }
\newcommand{\calh}{ {\cal H} }
\newcommand{\calo}{ {\cal O} }

\def \appb{{\bf Acta. Phys. Polon. B }  }
\def \cpc{ {\bf Chin. Phys. C } }
\def \ctp{ {\bf Commun. Theor. Phys. } }
\def \epjc{{\bf Eur. Phys. J. C} }
\def \ijmpcs{{\bf Int. J. Mod. Phys. Conf. Ser.} }
\def \jhep{{\bf J. High Energy Phys. } }
\def \jpg{ {\bf J. Phys. G} }
\def \mpla{{\bf Mod. Phys. Lett. A } }
\def \npb{ {\bf Nucl. Phys. B} }
\def \plb{ {\bf Phys. Lett. B} }
\def \ppn{ {\bf Phys. Part. Nucl. } }
\def \ppnp{{\bf Prog.Part. Nucl. Phys.  } }
\def \pr{  {\bf Phys. Rep.} }
\def \prc{ {\bf Phys. Rev. C }}
\def \prd{ {\bf Phys. Rev. D} }
\def \prl{ {\bf Phys. Rev. Lett.}  }
\def \ptp{ {\bf Prog. Theor. Phys. }}
\def \zpc{ {\bf Z. Phys. C}  }
\def \jpg{ {\bf J.Phys.-G-}  }
\def \ap{ {\bf Ann. of Phys}  }

\preprint{SI-HEP-2018-30}
\preprint{QFET-2018-19}

{\footnotesize

\title{${\bf Z \to \pi^+\pi^-, K^+K^-}$: A touchstone of the PQCD approach}

\author{Shan Cheng$^{a}$} \email{scheng@hnu.edu.cn}
\author{Qin Qin$^b$} \email{qin@physik.uni-siegen.de} 

\affiliation{$^a$ \, School of Physics and Electronics, Hunan University, 410082 Changsha, People's Republic of China, \non
$^b$ \, Theoretische Physik 1, Naturwissenschaftlich-Technische Fakult\"{a}t,  Universit\"{a}t Siegen, Walter-Flex-Strasse 3, D-57068 Siegen, Germany} 

\date{\today}
             
\begin{abstract}

We study two rare decays, $Z \to \pi^+\pi^-$ and $K^+K^-$, in the perturbative QCD approach 
up to the next-to-leading order of the strong coupling
and the leading power of $1/m_Z$, $m_Z$ being the $Z$ boson mass. The branching ratios 
$\mathcal{B}(Z\to \pi^+\pi^-) = (0.83 \pm 0.02 \pm 0.02 \pm 0.04)\times 10^{-12}$ and 
$\mathcal{B}(Z\to K^+K^-) = (1.74^{+0.03}_{-0.05} \pm 0.04 \pm 0.02)\times 10^{-12}$ are
obtained and can be measured at a tera-$Z$ factory. Because the 
subleading-power contributions to the branching ratios are negligible,
and the leading one does not depend on any free parameter, the two channels can 
serve as a touchstone for the applicability of 
the perturbative QCD approach.

\end{abstract}

\maketitle

\section{Introduction}

Two-body nonleptonic $B$ meson decays play an essential role in particle physics and help us understand the 
QCD and the charge conjugation parity violation in the Standard Model. They 
have inspired the development of many theoretical frameworks or approaches, 
including the QCD-improved factorization approach \cite{BenekeBR}, 
the soft-collinear effective theory \cite{BauerEW,Beneke:2002ph}, 
the light-cone sum rules \cite{KhodjamirianMI}, 
the perturbative QCD (PQCD) approach \cite{LiUN} based on the $k_T$ factorization theorem \cite{CataniXK,CollinsTY,BottsKF,HuangRD}, 
and the factorization-assisted topological-amplitude approach \cite{Li:2012cfa} proposed recently. 
Among them, the PQCD approach is the most predictive one, in which
a high-energy hadronic process is factorized into universal distribution amplitudes of hadrons 
and a perturbatively calculable hard kernel. 
However, it is also this unique feature of PQCD that has been questioned. 
The power counting analyses of the $B\to\pi$ form factor and the timelike pion form factor \cite{BenekeBR,BraunUJ} 
imply that both the nonperturbative small-$x$
($x$ is the momentum fraction of a constitute quark in a pion) region and the perturbative 
$x\sim1/2$ region contribute at the leading power of $1/m_B$, $m_B$ being the $B$ meson mass. 
On the other hand, a PQCD calculation shows that the small-$x$ region is practically suppressed by the 
Sudakov factor from the $k_T$ 
resummation, and thus the form factors are dominated by perturbative contributions \cite{StermanAJ,CataniNE,LiGI}. 
To test which argument is valid, we propose the $Z^0\to\pi^+\pi^-$ ($K^+K^-$) channel as a touchstone here. 
In the PQCD approach to the $Z\to\pi^+\pi^-$ decay rate, power corrections in $1/m_Z$, 
$m_Z$ being the $Z$ boson mass, are so small that they 
can be neglected safely. It hints that we need to consider only the twist-2 light-cone distribution amplitude 
(LCDA) of the pion, and that its simple asymptotic form may be justified. As a result, the calculation is free of 
arbitrariness, since the nonperturbative pion LCDA has been fixed \cite{FollanaUV,AokiFRL,GelhausenWIA}.
The two channels are expected to be observed or strictly constrained at a future tera-$Z$ factory like 
the FCC-ee, formerly known as TLEP \cite{Gomez-Ceballos:2013zzn}, and/or the Circular Electron-Positron 
Collider \cite{CEPCStudyGroup:2018rmc}, which can be used not only to precisely study the Higgs and $Z$ properties 
(\textit{e.g.} see \cite{QinAJU}) and discover new particles (\textit{e.g.} see \cite{AliIFM}), 
but also to improve our understanding of QCD as elaborated in this paper.

The $Z\to\pi^+\pi^-$ decay amplitude is proportional to the timelike pion form factor, which can be 
investigated in several different methods in principle. 
One is the partial wave analysis, in which elastic and inelastic scatterings as well as 
effects of resonances are handled \cite{RopertzSTK,DaiZTA}.
Another one, the light-cone sum-rule approach, is powerful for spacelike form factors, while 
dispersion relations and some resonance models are inevitable for the timelike region \cite{GeshkenbeinGU}.
Both the above approaches work well only in the low-energy region and are model dependent. 
To access the form factor with the dipion invariant mass at order of $m_Z$, the PQCD
approach is more appropriate \cite{HuCP,Gousset:1994yh,Qin:2015tba,ChenTCH}. 
In this paper, we will evaluate the $Z^0\to\pi^+\pi^-$ ($K^+K^-$) decay rate up to 
the next-to-leading order (NLO) of the strong coupling $\alpha_s$ and at the leading 
power of $1/m_Z$ in the PQCD formalism \cite{LiUN}. We obtain the branching ratio about 
0.83 (1.74) $\times 10^{-12}$, which is likely to be measured at a tera-$Z$ factory. 
Whichever of them is found, it will be the first observation of an exclusive hadronic 
$Z$ decay and serve as a touchstone to verify the PQCD approach. 

The rest of the paper is organized as follows. 
In Section \ref{sec:framework} the PQCD calculation of the timelike pion form factor is performed up to NLO, 
and the analytical formulas are given. 
In Section \ref{sec:numerics} we present the numerical results for the $Z \to \pi^+ \pi^-, K^+K^-$ branching ratios.
Section \ref{sec:conclusion} is the conclusion.

\section{Perturbative calculation}\label{sec:framework}

In the Standard Model, the $Z\bar{q}q$ interaction is described by $- J_\mu^{(Z)}Z^\mu$ in the Lagrangian with the current
\begin{align}
J_\mu^{(Z)} = {g\over2\cos\theta_w}\sum_{q}\left[(T_q-2Q_q\sin^2\theta_w)\bar{q}\gamma_\mu q - T_q\bar{q}\gamma_\mu\gamma_5q\right],
\end{align}
where $g$ is the SU(2) gauge coupling, $\theta_w$ is the weak mixing angle, and the hypercharges and electric charges 
of the quarks are $T_{u,d} = \pm1/2$, and $Q_u=$ 2/3 and $Q_d$ = -1/3, respectively. 
We then write the $Z\to \pi^+\pi^-$ decay amplitude
as 
\begin{align}
i\mathcal{M}(Z\to\pi^+\pi^-) =  \langle \pi^+\pi^- | J_\mu^{(Z)} | 0\rangle \epsilon_Z^\mu,
\end{align}
with the polarization vector $\epsilon_Z$ of the decaying $Z$ boson. 
Only the vector components in $J_\mu^{(Z)}$ contribute, because hadronic matrix elements 
induced by the axial-vector currents are forbidden by parity.
The timelike pion form factor $\mathcal{G}(Q^2)$ is defined via
\begin{align}
\langle \pi^+\pi^-|\bar{u}\gamma^\mu u | 0\rangle = (p_1^\mu - p_2^\mu)\mathcal{G}(Q^2), \qquad 
\langle \pi^+\pi^-|\bar{d}\gamma^\mu d | 0\rangle = -(p_1^\mu - p_2^\mu)\mathcal{G}(Q^2),
\end{align}
with $p_{1}$ and $p_{2}$ being the momenta of $\pi^+$ and $\pi^-$, respectively, $q=p_1+p_2$, and $Q^2 = q^2$.
The above two definitions are equivalent due to the isospin symmetry. 
Performing the phase space integral, we obtain the spin-averaged decay width 
\begin{align}
\Gamma(Z\to\pi^+\pi^-) &={1\over3}{1\over16\pi m_Z}\sum_{s}\left|\mathcal{M}(Z\to\pi^+\pi^-)\right|^2, \nonumber\\
&={1\over3}{1\over16\pi m_Z}(g_V^u-g_V^d)^2\left|\mathcal{G}(m_Z^2)\right|^2(p_1^\mu - p_2^\mu)(p_1^\nu - p_2^\nu)\sum_s\epsilon^*_\mu(P_Z)\epsilon_\nu(P_Z), \nonumber\\
&={m_Z\over48\pi}(g_V^u-g_V^d)^2\left|\mathcal{G}(m_Z^2)\right|^2,
\label{eq:width}
\end{align}
where $g_V^q ={g/(2\cos\theta_w)}\times(T_q-2Q_q\sin^2\theta_w)$ \cite{FanchiottiTU,MarcianoDP}, 
and the pion mass effect has been neglected. 
The factor $(g_V^q-g_V^{q^\prime})^2$ indicates that the $Z\to\pi^0\pi^0$ and $K^0\bar{K}^0$ 
decays are forbidden at leading power. Below, we focus on the evaluation of the form factor 
$\mathcal{G}(Q^2)$ at $Q^2=m_Z^2$ in the PQCD approach.


\subsection{Kinematics and the LO form factor}

\begin{figure}[tb]
\begin{center}
\vspace{-1cm}
\includegraphics[width=0.6\textwidth]{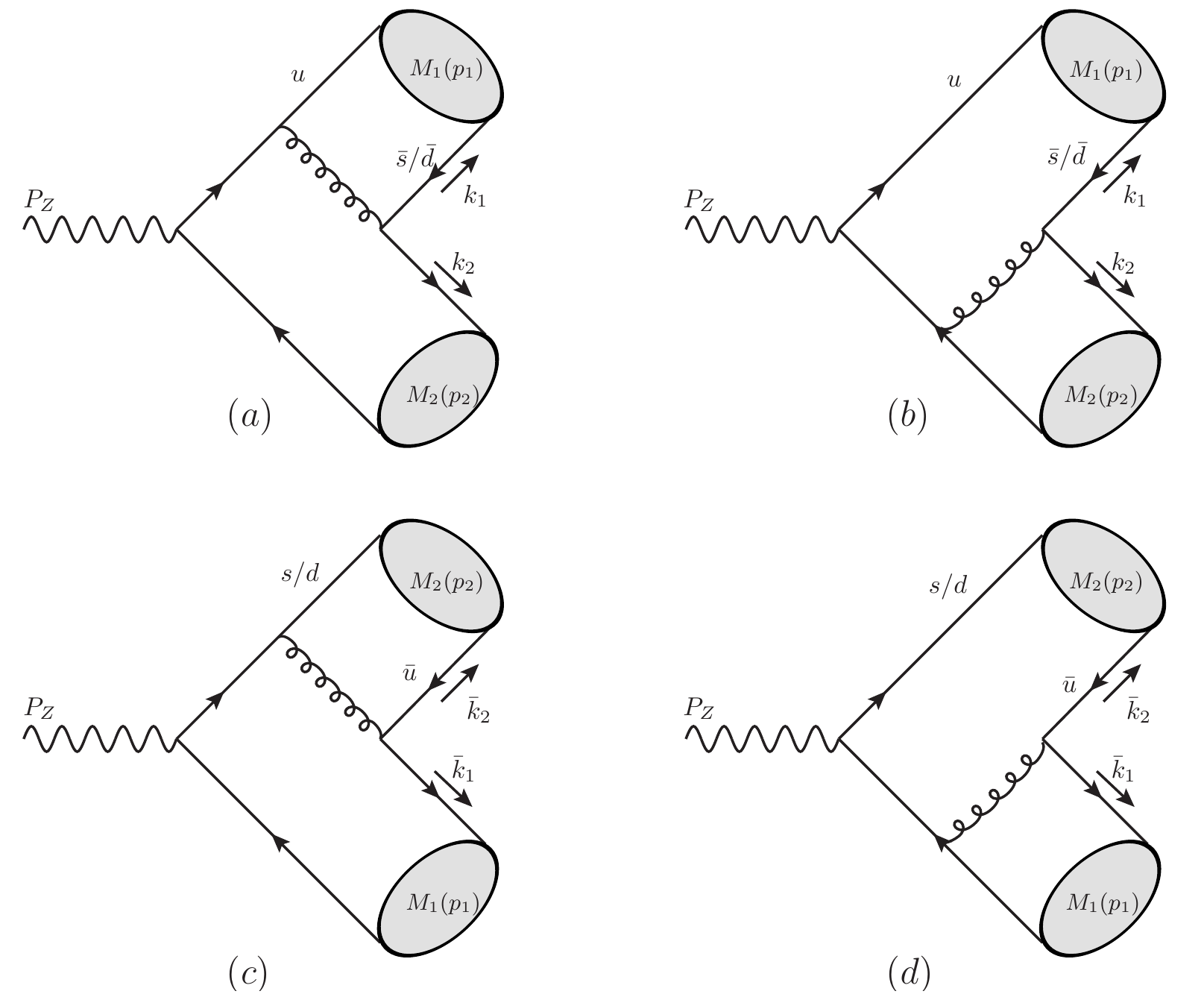}
\end{center}
\vspace{-1cm}
\caption{\footnotesize{Feynman diagrams for $Z \to M_1M_2$ decays at leading order
with $M_1= \pi^+, K^+$ and $M_2 = \pi^-, K^-$.}}
\label{fig:1}
\end{figure}

As depicted in Fig. \ref{fig:1}, the two upper and lower diagrams contribute to the timelike form factors from 
$\langle \pi^+\pi^-|\bar{u}\gamma^\mu u | 0\rangle$ and $\langle \pi^+\pi^-|\bar{d}\gamma^\mu d | 0\rangle$, 
respectively, at leading order (LO) of QCD. We choose the following kinematics for the initial- and the final-state 
particles expressed in terms of light-cone coordinates, 
\beq
p_Z = \frac{m_Z}{\sqrt{2}} (1, 1, {\bf 0}),\qquad p_1 = \frac{m_Z}{\sqrt{2}}(1, 0, {\bf 0}),\qquad p_2 =  \frac{m_Z}{\sqrt{2}}(0, 1, {\bf 0}),
\eeq
where $p_Z$ is the momentum of the $Z$ boson.
The $Z$ boson is at rest in this frame, and the two pion momenta are collimated to the two light-cone 
directions, with the pion masses being ignored. 
The momenta of the constitute quarks and antiquarks in Fig. \ref{fig:1} are parametrized as 
\beq
{\footnotesize
k_1 =  (x_1\frac{m_Z}{\sqrt{2}}, 0, {\bf k}_{1T}),\qquad
k_2 = ( 0, x_2\frac{m_Z}{\sqrt{2}}, {\bf k}_{2T}),\qquad
\bar{k}_1 = p_1-k_1,\qquad \bar{k}_2=p_2-k_2.
}\eeq
We can get the pion form factor at leading power 
\footnote{Details of the calculation and the factorization formula for the contribution
from higher-twist LCDAs are given in Appendix. \ref{app:ff}.}  
by computing any diagram in Fig. \ref{fig:1}, 
\beq
\mathcal{G}_{\text{II}}(Q^2)_\text{LO} =& \ -{16 \pi C_F} Q^2 \int_0^1 dx_1dx_2 \int db_1 db_2 b_1 b_2 \, \alpha_s(\mu) x_2 \phi_\pi(x_1) \phi_\pi(x_2) \nonumber\\
&\times h_\text{II}(x_1, b_1,x_2,b_2,Q) \, \mathrm{Exp}[-S_\text{II}(x_1,b_1,x_2,b_2,\mu)] \,,
\label{eq:ff-lo} 
\eeq
where $C_F =4/3$, $b_{i}$ are the conjugate variables of the transverse momenta ${\bf {k}}_{iT}$, 
and $\phi_\pi(x)$ is the twist-2 pion LCDA. The factorization scale $\mu$ is set to $\max(1/b_1,1/b_2,\sqrt{x_2}Q)$.
The Sudakov factor derived from the $k_T$ resummation up to the next-to-leading-logarithm 
accuracy is written as
\beq
{\footnotesize S_\text{II}(x_i,b_i,\mu)  = \sum_{i=1,2} \left[ s(x_i\frac{m_Z}{\sqrt{2}}, b_i ) + s ((1-x_i) \frac{m_Z}{\sqrt{2}}, b_i ) + s_q (b_i, \mu)  \right] \,,}
\eeq
where the terms $s(Q_i, b_i)$ collect the double and single logarithms in the 
vertex correction associated with an energetic light quark (see Eq. (10) of \cite{BottsKF}), 
and the term
\beq
s_q(b, \mu)\ &&= -2\int_{1/b}^\mu{d\bar{\mu}\over\bar{\mu}}{\alpha_s(\bar{\mu})\over\pi}\non
&&= - {1\over\beta_1}\log\left( {\log(t/\Lambda^{(5)})\over -\log(b\Lambda^{(5)})} \right)
- {\beta_2\over 2\beta_1^3} \left( {\log[2\log(t/\Lambda^{(5)})]+1 \over \log(t/\Lambda^{(5)})} - {\log[-2\log(b\Lambda^{(5)})] +1\over -\log(b\Lambda^{(5)}))}  \right),
\eeq
resums the single logarithms in the quark self-energy correction \cite{LiUN}. 
We adopt the two-loop expression for the strong coupling 
\beq
\alpha_s(\mu)\ &&= {\pi\over 2\beta_1\log\left(\mu/\Lambda^{(5)}\right)}\left(1 - {\beta_2\over \beta_1^2}
{\log(2\log(\mu/\Lambda^{(5)}))\over 2\log(\mu/\Lambda^{(5)})} \right),
\eeq
with $\beta_1 = (33-2n_f)/12$, $\beta_2 = (153-19n_f)/24$ and the flavor number $n_f =5$. 
We do not take into account the threshold resummation factor \cite{StermanAJ,CataniNE,LiGI} for the hard kernel, 
which is important only for subleading contributions from higher-twist LCDAs.

The hard function $h_\text{II}(x_1, b_1,x_2,b_2,Q)$ in the form factor contains 
the internal propagators expressed in the coordinate 
space conjugate to the transverse momenta:
\beq
&&\int{d^2{\bf b}_1d^2{\bf k}_{1T}\over(2\pi)^2} \int{d^2{\bf b}_2d^2{\bf k}_{2T}\over(2\pi)^2} 
e^{-i({\bf k}_{1T}\cdot {\bf b}_{1} + {\bf k}_{2T}\cdot {\bf b}_{2})} 
{1\over x_2Q^2 -{\bf k}_{2T}^2 +i\epsilon}{1\over x_1x_2Q^2 -({\bf k}_{1T}+{\bf k}_{2T})^2 +i\epsilon} \nonumber \\
= &&\int_0^\infty db_1db_2b_1b_2\left({i\pi\over2}\right)^2H^{(1)}_0(\sqrt{x_1x_2}Q b_1)[\theta(b_1-b_2)J_0(\sqrt{x_2}Q b_2)H^{(1)}_0(\sqrt{x_2}Q b_1) 
+ (b_1\leftrightarrow b_2)] \non
\equiv && \int_0^\infty db_1db_2b_1b_2\ h_\text{II}(x_1,b_1,x_2,b_2,Q),
\label{eq:bessel-ii}
\eeq
in which $J_0$ is the Bessel function of the first kind and $H^{(1)}_0$ is the Hankel function of the first kind.
We notice that Eq \eqref{eq:bessel-ii} oscillates violently as $Q^2$ goes beyond 50 $\mathrm{GeV}^2$, 
resulting from the large hierarchy between the two scales, $Q^2$ and $k_{T}^2$. The strong oscillation 
causes difficulty in obtaining the convergent multiple integral in \eqref{eq:ff-lo} numerically.
\footnote{This hierarchy is less obvious in $B$ meson decays because of $Q^2 = m_B^2$, 
and the numerical integrals converge quickly.}
To overcome this difficulty, we assume the hierarchy ansatz $x_i Q^2 \gg x_1x_2 Q^2 \sim k_T^2$ 
according to the power counting in the PQCD approach, 
dropping the transverse momentum in the quark propagator but retaining the transverse momentum 
in the propagator of the hard gluon. 
As a consequence, the double-$b$ hard function in \eqref{eq:bessel-ii} is reduced to a single-$b$ one, 
\beq
&&\int{d^2{\bf b}_1d^2{\bf k}_{1T}\over(2\pi)^2} \int{d^2{\bf b}_2d^2{\bf k}_{2T}\over(2\pi)^2} 
e^{-i({\bf k}_{1T}\cdot {\bf b}_{1} + {\bf k}_{2T}\cdot {\bf b}_{2})} 
{1\over x_2Q^2 +i\epsilon}{1\over x_1x_2Q^2 -({\bf k}_{1T}+{\bf k}_{2T})^2 +i\epsilon} \nonumber \\
= &&\int_0^\infty dbb {1\over x_2Q^2}\left(-{i\pi\over2}\right)H^{(1)}_0(\sqrt{x_1x_2}Q b) \non
\equiv && \int_0^\infty dbb \ h_\text{I}(x_1,x_2,b,Q) \,,
\eeq
with $b = b _1 = b_2$ read off the above derivation. 
This approximation simplifies the computational task and also extends the numerically
manageable range in $Q^2$ from dozens to thousands of $\mathrm{GeV}^2$. 
The form factor at LO is then modified to 
\beq
\mathcal{G}_{\text{I}}(Q^2)_\text{LO} &=& -{16 \pi C_F} Q^2\int_0^1 dx_1dx_2 \int db b \, \alpha_s(\mu)x_2\phi_\pi(x_1) \phi_\pi(x_2) 
h_\text{I}(x_1,x_2, b, Q) \, \mathrm{Exp}[-S_\text{I}(x_1,x_2,b,\mu)] \non
&=& i{8 \pi^2 C_F}\int_0^1 dx_1dx_2 \int db b \, \alpha_s(\mu)\phi_\pi(x_1) \phi_\pi(x_2) 
H^{(1)}_0(\sqrt{x_1x_2}Q b) \, \mathrm{Exp}[-S_\text{I}(x_1,x_2,b,\mu)] \, ,
\label{eq:ff-lo-singleb} 
\eeq
with $S_\text{I}(x_1,x_2,b,\mu) = S_\text{II}(x_1,b,x_2,b,\mu)$. As will be observed in 
Fig. \ref{fig:2}, in which the double-$b$ and single-$b$ results are compared, the single-$b$ approximation 
works very well in the high-$Q^2$ region.

\subsection{Next-to-leading-order QCD correction}

The NLO correction to the timelike pion form factor has been explored in the PQCD approach with the 
single-$b$ convolution \cite{Li:2010nn},
\begin{align}
\mathcal{G}_{\text{I}}(Q^2)_\text{NLO} =&\ i{2 \pi C_F^2} \, \int_0^1 dx_1dx_2 \int db b \alpha^2_s(\mu) \phi_\pi(x_1) \phi_\pi(x_2) 
\mathrm{Exp}[-S_\text{I}(x_1,x_2,b,\mu)] \,, \non
&\times \left[ \tilde{h}(x_1,x_2,b,Q,\mu)H^{(1)}_0(\sqrt{x_1x_2}Qb) + H^{(1)\prime\prime}_0(\sqrt{x_1x_2}Qb) \right] \, ,
\label{eq:ff-nlo-singleb} 
\end{align}
where the explicit expression of the NLO function $\tilde{h}(x_1,x_2,b,Q,\mu)$ is referred to Eq. (18) of Ref. \cite{HuCP}. 
For the second derivative of the Hankel function on the order parameter
\beq
H^{(1)\prime\prime}_0(x) \equiv \left[ {d^2\over d\alpha^2}H^{(1)}_\alpha(x)\right]_{\alpha=0}, 
\eeq
we take the following fit function in practice
\beq
\text{Re}[H^{(1)\prime\prime}_0(x)] &=&
\left\{ \begin{array}{ll} 
0.798 + 0.454 x -0.0603 x^2 +0.00590 x^3 -0.00021 x^4  -1.35 \log{x}  & \\ 
+ J_0(x) \left(-0.581 +1.48 \log{x} - 0.497 \log^2{x} \right) &  \\
+ Y_0(x) \left(-3.62 -0.194 x + 0.665 \log{x} + 0.331 \log^2{x}\right),& x \geqslant 10, \\  
 & \\[-0.4cm]
 \left(-0.0870 x^{-5/2} + 1.05 x^{-3/2}\right) \cos{(\frac{\pi}{4}-x)} & \\ 
+ \left(-0.624 x^{-5/2} -0.000588 x^{-3/2} +1.97 x^{-1/2} \right) \sin{(\frac{\pi}{4}-x)}, & x < 10, \\ 
\end{array}\right. \label{eq:Hpp-re}\\
\text{Im}[H^{(1)\prime\prime}_0(x)] &=& 
\left\{ \begin{array}{ll} 
-4.58 + 0.720 x -0.151 x^2 +0.00643 x^3 + 2.57 \log{x} & \\ 
+ J_0(x) \left(3.16 - 0.794 x + 0.0179 x^2 - 5.65 \log{x} + 2.26 \log^2{x} \right) &  \\
+ Y_0(x) \left(4.10 - 2.03 \log{x} -0.00708 \log^2{x}\right) ,& x \geqslant 10, \\  
 & \\[-0.4cm]
\left(0.610 x^{-5/2} + 0.00182 x^{-3/2} -1.97 x^{-1/2} \right) \cos{(\frac{\pi}{4}-x)} & \\ 
+ \left(-0.0897 x^{-5/2} +1.05 x^{-3/2} \right) \sin{(\frac{\pi}{4}-x)}, & x < 10. \\ 
\end{array}\right. 
\label{eq:Hpp-im}
\eeq 
The NLO correction \eqref{eq:ff-nlo-singleb} was applied to analyze the $B_c$ pair production 
at electron-positron colliders with the nonrelativistic-QCD $B_c$ meson distribution amplitudes 
\cite{WeiXLR} recently, in which only the small argument limit for $H_0^{(1) \prime \prime}(x)$ 
was considered.

\section{Numerics}\label{sec:numerics}

The asymptotic form of the twist-2 pion LCDA is employed here, 
\beq
{\footnotesize
\phi_\pi(x) = \frac{f_\pi}{2\sqrt{2N_c}} 6 x(1-x) \,,
}\eeq
with the pion decay constant $ f_\pi = 130.2\pm1.4 \mathrm{GeV} $ \cite{AokiFRL}. 
The other numerical inputs include \cite{Tanabashi:2018oca} the width $\Gamma_Z = 2.4952 \pm 0.0023$ GeV and  
\beq
&&\sin^2 \theta_w \, (m_Z) = 0.23129\pm 0.00005 \,, \,\,\, \alpha_s(m_Z) = 0.1182\pm0.0012 , \,\,\,
\alpha(m_Z)^{-1} = 127.950 \pm 0.017 , 
\eeq
defined at the $m_Z$ scale under the modified minimal subtraction ($\overline{\text{MS}}$) scheme.
To reproduce the central value of $\alpha_s(m_Z)$ with the two-loop accuracy, the scale 
$\Lambda^{(5)}_{\overline{\text{MS}}} = 0.2327 \, \mathrm{GeV}$ is chosen.
Using a Monte-Carlo integration strategy with the Vegas \cite{LepageDQ} algorithm from the GNU Scientific 
Library \cite{GSL:Vegas}, we estimate the integral with 500,000,000 sampling points for the real and 
imaginary parts of Eqs \eqref{eq:ff-lo-singleb} and 
\eqref{eq:ff-nlo-singleb}, which achieves a relative precision better than permillage level. 
The central values of the LO pion form factor 
and the NLO correction at $Q^2=m_Z^2$ are 
\beq
&&\mathcal{G}(m_Z^2)_\text{LO}=(- 8.29  - i \, 0.771) \times10^{-6}  \,,\qquad \mathcal{G}(m_Z^2)_\text{NLO} = (-0.764 - i \, 1.58) \times 10^{-6}\,, 
\eeq
from which we see that the NLO correction enhances the LO result reasonably by about 10\%. 
The PQCD prediction up to NLO for the $Z\to \pi^+\pi^-$ branching ratio is given by
\beq
\mathcal{B}(Z\to \pi^+\pi^-) = (0.83 \pm 0.02 \pm 0.02 \pm 0.04)\times 10^{-12},
\eeq
with the three uncertainties coming from the scale variation from $\mu/2$ to 2$\mu$, the strong coupling constant and the pion 
decay constant, respectively. Replacing the pion decay constant in the calculation with the kaon decay constant 
$f_K = 155.6\pm0.4 \, \mathrm{GeV} $ \cite{AokiFRL}, we have
the corresponding $Z\to K^+K^-$ branching ratio
\beq
\mathcal{B}(Z\to K^+K^-) = (1.74^{+0.03}_{-0.05} \pm 0.04 \pm 0.02)\times 10^{-12}.
\eeq
According to \cite{CEPCStudyGroup:2018rmc}, the Circular Electron-Positron Collider is expected 
to collect $7\times10^{11}$ $Z^0$ bosons in two years with the instantaneous luminosity of 
$32\times10^{34}$ cm$^{-2}$s$^{-1}$ and two interaction points. The FCC-ee 
\cite{Gomez-Ceballos:2013zzn}, with the instantaneous luminosity of 
$56\times10^{34}$ cm$^{-2}$s$^{-1}$ and four interaction points, will 
quadruple this number roughly. If the two 
channels are combined, observations at the two tera-$Z$ factories will be 
quite promising, owing to almost 100\% detection 
efficiencies of charged pions and kaons. On the other hand, if the contributions from the "small-$x$" region 
are actually dominant in the pion and kaon timelike form factors as postulated in \cite{BenekeBR,BraunUJ}, 
more events will be expected. 

To confirm the validity of the single-$b$ configuration, we compute the LO timelike pion form factor 
in the region $Q^2 \in [1, 50] \, \mathrm{GeV}^2$ using both the double-$b$ and single-$b$ formulas \eqref{eq:ff-lo} 
and \eqref{eq:ff-lo-singleb}, and display them in Fig. \ref{fig:2}. The discrepancy between the two results is visible 
in the low-$Q^2$ region, while starting from $\sim$ 40 GeV$^2$, we can safely omit the transverse momentum
effect in the internal quark propagator and adopt the single-$b$ approximation.

\begin{figure}[!tb]
\begin{center}
\vspace{-1cm}
\includegraphics[width=0.415\textwidth]{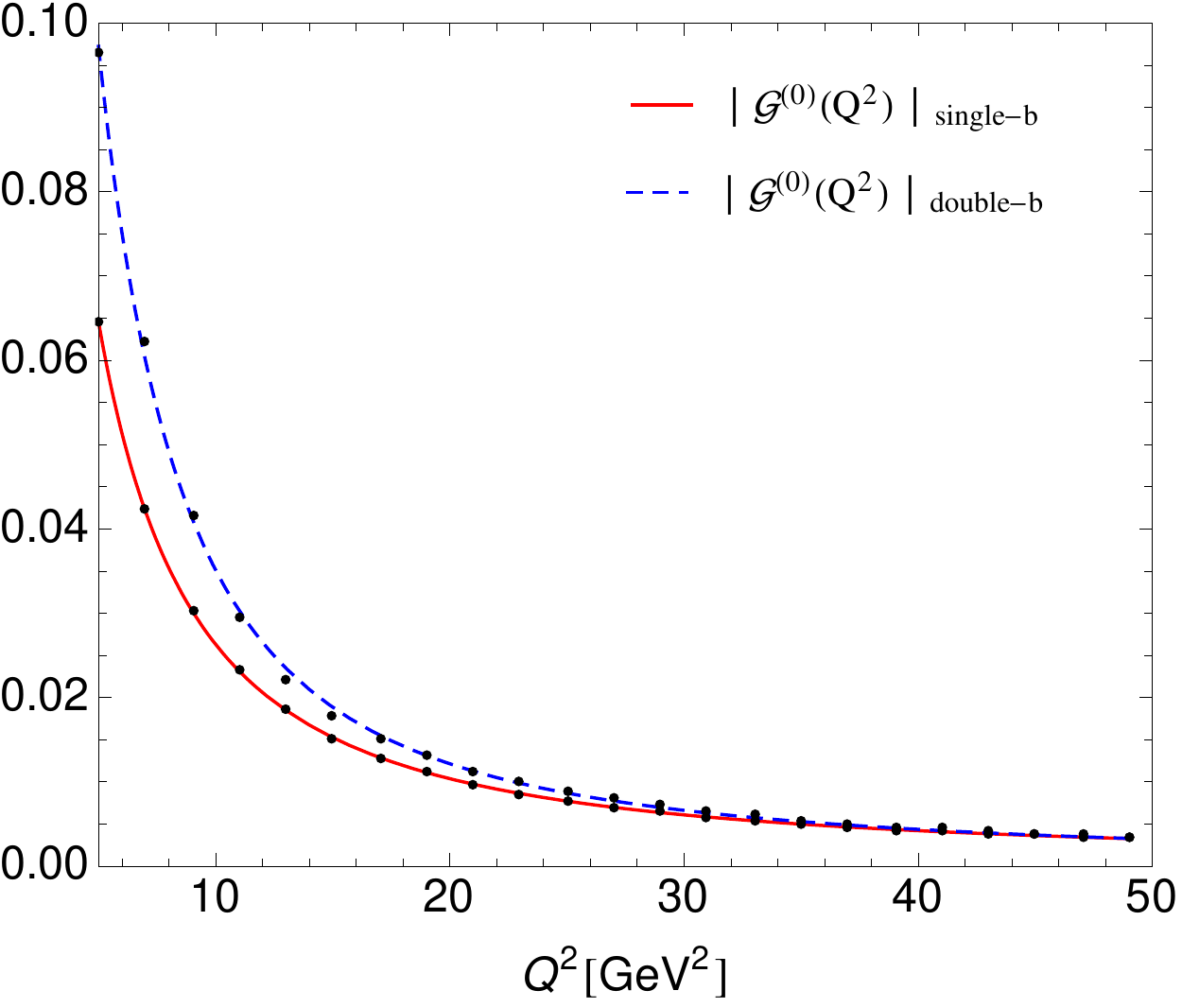}
\end{center}
\vspace{-1cm}
\caption{{\footnotesize Magnitude of the LO timelike pion form factor $\mathcal{G}(Q^2)$ 
derived in the double-$b$ and single-$b$ formulations.}}
\label{fig:2}
\end{figure}

The magnitude and the strong phase of the pion form factor at LO and NLO for $Q^2$ between 
50 GeV$^2$ and $m_Z^2$ are shown in Fig. \ref{fig:3}.
The NLO correction to the magnitude is found to be around $11 \%$ in the whole considered
$Q^2$ range. For the strong phase, the LO prediction is about 180$\,^{\circ}$, and the 
NLO correction yields an increase not more than 20$\,^{\circ}$
\footnote{The NLO correction brings a large enhancement to the imaginary part, but it is
still considerably smaller than the LO real part.}.
We suggest a parametrization formula for the form factor far away from the 
resonance region with $Q$ in units of GeV, 
\beq
{\footnotesize
\left| \mathcal{G} (Q^2)\right| = \frac{A + Q^2 B}{Q^4 + Q^2 C + A} \, ,
\label{eq:para}
}\eeq
which is inspired by the parametrization with the reciprocal of the square 
polynomial \cite{KhodjamirianTK}. Here we have added another $Q^2$ term in 
the numerator to relieve a sudden drop at $Q^2$ around several hundred $\mathrm{GeV}^2$.
The equality of the constant terms in the numerator and the denominator is motivated 
by the normalization condition of the pion form factor 
$\mathcal{G}_\pi(0) = 1$ (for references, see \textit{e.g.} \cite{Cheng:2017smj}).
For the LO timelike pion form factor, the parameters $A^{(0)} = 0.0879, B^{(0)}= 46.1$, and 
$C^{(0)}=10.9$ are determined.
Including the NLO correction, we have $A= 0.0996, B= 48.2$, and $C=12.6$. 
\begin{figure}[tb]
\begin{center}
\vspace{-1cm}
\includegraphics[width=0.465\textwidth]{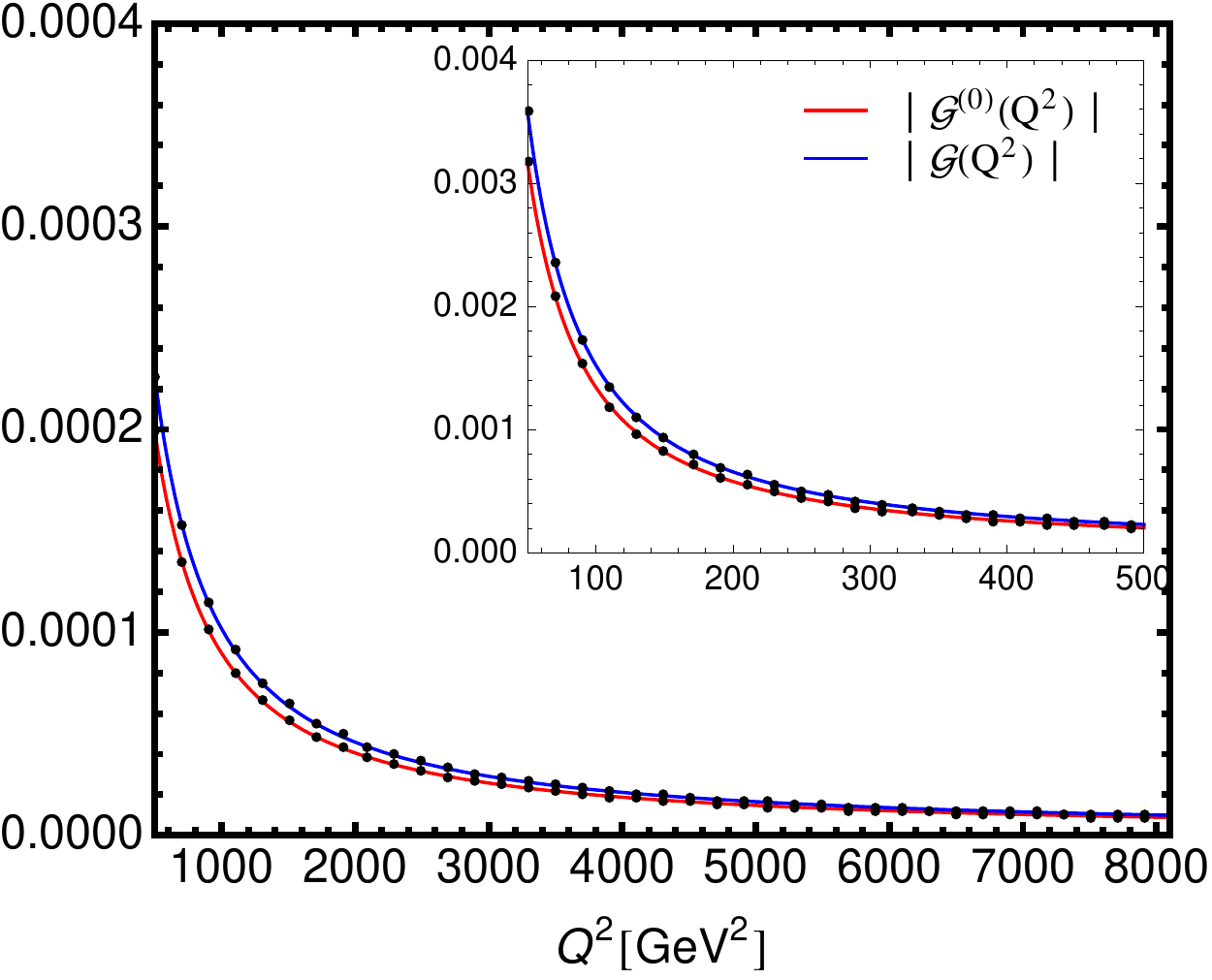}
\hspace{4mm}
\includegraphics[width=0.44\textwidth]{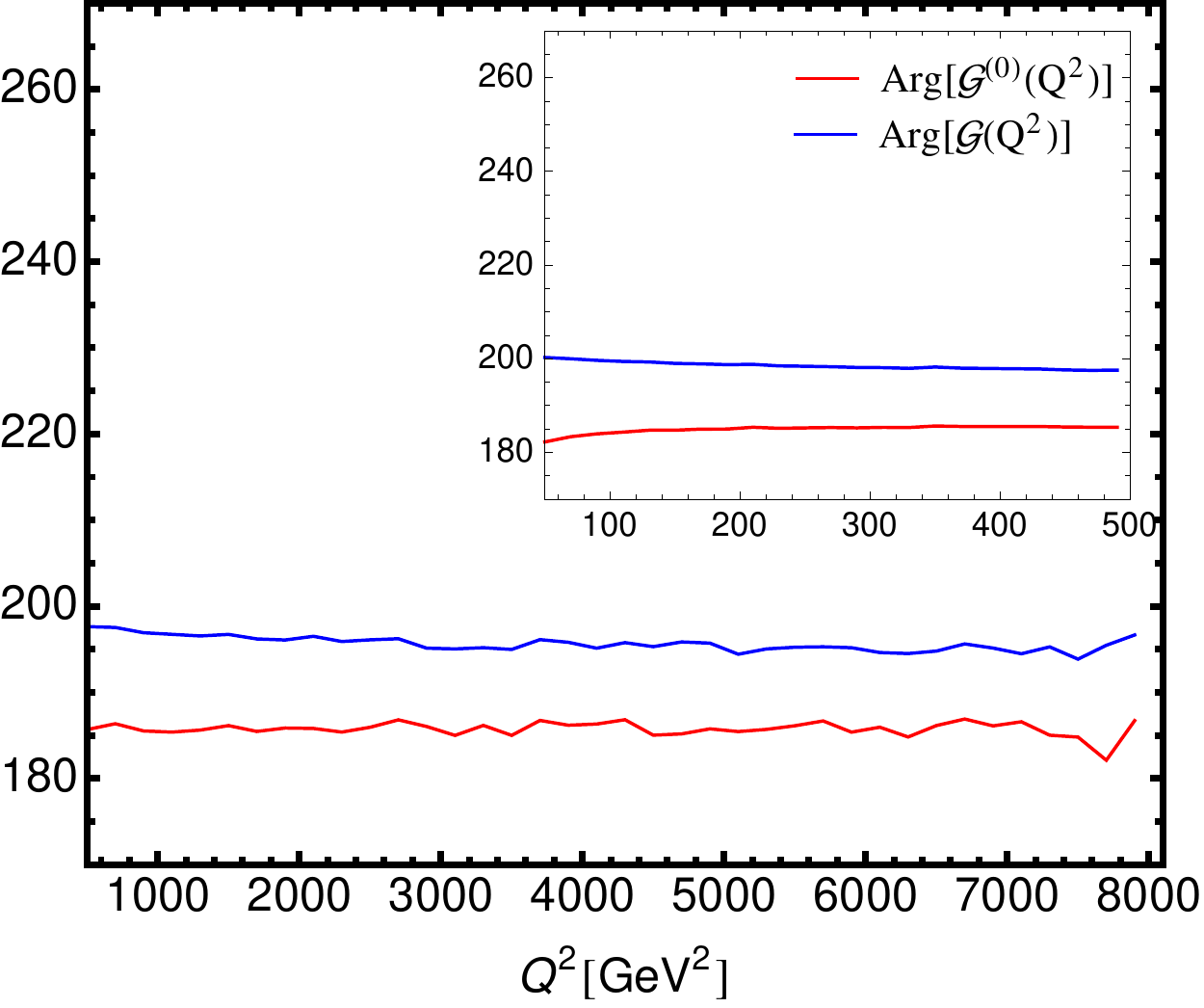}
\end{center}
\vspace{-1cm}
\caption{{\footnotesize LO ($\mathcal{G}^{(0)}$) and NLO ($\mathcal{G}$) predictions for the magnitude (left) and the phase (right) of the pion form factor for $Q^2$ between 50 GeV$^2$ and $m_Z^2$.} }
\label{fig:3}
\end{figure}

\section{Conclusion}\label{sec:conclusion}

We have studied the $Z \to \pi^+\pi^-, K^+K^-$ decays in the PQCD formalism, whose branching 
ratios are governed by the timelike form factors of the corresponding mesons. 
With a high $Q^2$ = $m_Z^2$, we can safely neglect the power corrections in the 
PQCD evaluation of the form factors, 
which then do not depend on any unknown nonperturbative parameters and can be
predicted precisely. Our predictions 
up to NLO for the the branching ratios of the two channels are 
$\mathcal{B}(Z\to \pi^+\pi^-) = (0.83 \pm 0.02 \pm 0.02 \pm 0.04)\times 10^{-12}$ and 
$\mathcal{B}(Z\to K^+K^-) = (1.74^{+0.03}_{-0.05} \pm 0.04 \pm 0.02)\times 10^{-12}$.
They can be accessed at a future tera-$Z$ factory, and the measurements will
represent a touchstone of the PQCD approach.

\section{ACKNOWLEDGEMENTS}

We are grateful to Hsiang-nan Li, Xin Liu, Yue-Long Shen and Yan-Bing Wei for helpful discussions, 
and especially to Hsiang-nan Li for the English language revision. 
S. C is supported by the National Science Foundation of China under No. 11805060 
and "the Fundamental Research Funds for the Central Universities" under No 020400/531107051171. 
Q. Q is supported by the DFG Research Unit FOR 1873 ``Quark Flavour Physics and Effective Theories''.  
S. C is grateful to Theoretical Division of Institute of High Energy Physics at Beijing for hospitality and for financial support where this work is finalized.

\begin{appendix}
\section{Pion form factor up to sub-leading twist }\label{app:ff}

The pion transverse-momentum-dependent wave function has been proposed 
in \cite{Collins:2011zzd} and \cite{Li:2014xda}, which
regularizes both the rapidity and self-energy divergences. Compared to 
\cite{Collins:2011zzd}, the form in \cite{Li:2014xda} is
simpler and compatible with the $k_T$ factorization. 
In the limit of vanishing infrared regulators, they both approach 
to the naive definition in \cite{Li:2013xna}. Here we assume that the dependence 
on the parton transverse momentum has been organized into the
Sudakov factor, and consider only the dependence on the longitudinal
momentum fraction, which can be formulated as \cite{BallWN}, 
\begin{align}
\langle  \pi^-(p_2) \vert \bar{d}_j(z) u(0)_l \vert 0 \rangle = \frac{i}{\sqrt{2N_c}} \int_0^1 e^{ix_2p_2 \cdot z} \gamma_5 \left\{
\psl_2 \phi_\pi(x_2) + m_0^\pi \phi_\pi^P(x_2) + m_0^\pi (\vsl \nsl -1) \phi_\pi^T(x_2) \right\}_{lj} \,.
\end{align}
Keeping the chiral mass $m_0^\pi$ which is expected to contribute the dominant subleading power correction, we find that 
the LO result of the pion form factor \eqref{eq:ff-lo} is modified to 
\beq
\mathcal{G}(Q^2) = &&\ -16 \pi C_F  Q^2\int_0^1 dx_1 dx_2 \int b_1db_1 b_2db_2 \, \alpha_s(\mu) \, h_\text{II}(x_1, b_1,x_2,b_2,Q) \, \mathrm{Exp}[-S_\text{II}(x_1,b_1,x_2,b_2,\mu)] \, \non
&& \times \left\{ x_2 \phi_\pi(x_1) \phi_\pi(x_2) + 2 r_\pi^2  \phi_\pi^P(x_1) \left( \phi_\pi^P(x_2) - \phi_\pi^T(x_2) \right)
+ 2 x_2 r_\pi^2 \phi_\pi^P(x_1) \left(\phi_\pi^P(x_2) + \phi_\pi^T(x_2) \right) \right\} \,,
\label{eq:ff-23}
\eeq 
with $r_\pi \equiv m_0^\pi/Q$. It is easy to confirm that the relative size of the power 
correction is of order of $r_\pi^2$ $\sim$ $10^{-4}$ at $Q = m_Z$.

\end{appendix}

}


\end{document}